# Limit of the Solutions for the Finite Horizon Problems as the Optimal Solution to the Infinite Horizon Optimization Problems

Dapeng CAI [1,*] and Takashi Gyoshin NITTA [2]

[1] *Institute for Advanced Research, Nagoya University, Furo-cho, Chikusa-ku, Nagoya, 464-8601, Japan;*  [2] *Department of Mathematics, Faculty of Education, Mie University, Kurimamachiya 1577, Tsu, 514-8507, Japan*

**Abstract**

We aim to generalize Cai and Nitta's (2007) results by allowing both the utility and production function to depend on time. We also consider an additional intertemporal optimality criterion. We clarify the conditions under which the limit of the solutions for the finite horizon problems is optimal among all attainable paths for the infinite horizon problems under the overtaking criterion, as well as the conditions under which such a limit is the unique optimum under the sum-of-utilities criterion. The results are applied to a parametric example of the one-sector growth model to examine the impacts of discounting on optimal paths.

---

[*] *Corresponding Author*. Tel/Fax: +81-52-788-6141; E-mail: cai@iar.nagoya-u.ac.jp.

## 1. Introduction

For growth models with infinite planning horizons, the infinite series of utility sequences may diverge, rendering the maximization of which meaningless in the real number field. The discounting approach, widely used in the field of intertemporal economics, avoids this problem as the discounted integral of utilities becomes finite under the "relentless force of compound interest" (Weitzman, 1998, p. 202). However, economists have long been scathing about the ethical dimensions and logical difficulties of discounting (Ramsey, 1928; Pigou, 1932; Harrod, 1948; Solow, 1974; Cline, 1992; Heal, 1998; Solow, 1998; Weitzman, 1998; Anannd and Sen, 2000).[1]

There has been a large literature that analyzes the undiscounted growth models. In Ramsey's original work, he formulated the problem as minimizing the deviation from a given reference curve, the "bliss level," which avoids the problem that the sum of the objective function may not converge. A more general approach was later introduced by Von Weizsächer (1965) and Atsumi (1965) and further refined by Gale (1967) and Brock (1970). This approach uses the concept of the overtaking criterion, which replaces the comparison of infinite sums by a comparison of finite partial sums. Under the assumption of the existence of such a given reference curve, the existence and dynamical properties of the resultant optimal path have been considered in, for example, Michel (1990), Durán (2000), Kamihigashi (2001), and Le Van and Morhaim (2006). Nevertheless, few studies have proposed how to explicitly construct the optimal paths for the undiscounted infinite horizon problems. Gale (1967) could be considered as a basis for such an approach. However, it would first involve checking, together with other assumptions, the existence of an optimal stationary path, which may be impossible for most problems.

---

[1] Heal (1998) asserts that "discounted utilitarianism has dominated until now largely because of the lack of a convincing alternative, not because of universal conviction about its merits" (p. 59). Stern (2007) concludes that "we [should] treat the welfare of future generations on a par with our own" (p. 35).

In Cai and Nitta (2007), a new approach that explicitly construct the optimal paths for the discrete time undiscounted infinite horizon optimization problems was proposed.[2] This approach does not require the assumption of the existence of an optimal path. They conclude that under a fairly general condition, the conjecture that "the limit of the solutions for the finite problems is optimal for the infinite horizon problem" is correct for the undiscounted problem, in the sense of the overtaking criterion.

The conjecture has been around for a while. The case when the discount factor is less than one has been examined by using recursive methods in, for example, Stokey and Lucas (1989). Proving the conjecture involves establishing the legitimacy of interchanging the operators "max" and "$\lim_{T \to \infty}$", which has been "more challenging than one might guess" for the discounted case (Stokey and Lucas, 1989, p. 12). Cai and Nitta (2007) has proved that the conjecture is correct for the undiscounted case, under certain conditions.

---

[2] By resorting to the knowledge of the non-standard analysis, they are able to specify the relevant boundary condition at the infinite terminal time as a natural extension of the boundary condition in the finite horizon case. They established the legitimacy of the conjecture by showing that the limit of the solutions for the finite problems is the unique "projection" of a unique hyper-real optimal path for the infinite horizon problem on the real number field.

However, Cai and Nitta (2007)'s analysis focuses exclusively on a particular example of the undiscounted infinite horizon problems, the approach is not directly applicable to other infinite horizon optimization problems. This paper aims at improving the applicability of their approach by generalizing their results. We allow both the utility function and the production function to depend on time, as in the standard theory of the calculus of variations. Our proof of the conjecture and the analysis are then extended to cover the discounted utility case. Hence, we have provided a full proof for the conjecture, covering both the undiscounted and the discounted cases.

Moreover, besides the overtaking criterion, we also consider another intertemporal optimality criterion: the sum-of-utilities criterion. We clarify the conditions under which the limits of the solutions for the finite horizon optimization problem is optimal among all attainable paths for the infinite horizon problem in the sense of the overtaking criterion, as well as the conditions under which such a limit is unique optimal under the sum-of-utilities criterion. The applicability of the new results is demonstrated by considering a parametric example.

The rest of the paper is organized as follows. In Section 2 we generalize Cai and Nitta's (2007) result. We clarify the conditions under which the conjecture holds under two criteria, respectively. In Section 3, we proceed to apply the results to a particular parametric example of the one-sector growth model and consider the impacts of discounting on the optimal capital accumulation paths. The legitimacy of the approach is discussed in Section 4. We then summarize and conclude. All proofs are relegated to the Appendix.

## 2. The Model and Results

### 2.1 The Setting

Cai and Nitta (2007) examine the conjecture that "the limit of the solutions for the finite problems is optimal for the infinite horizon problem" by considering stationary utility and production functions. We first generalize their results to a problem in which both the utility and production functions are allowed to depend on time. As argued in McKenzie (1986), the dependence on time "may take account of trends in technology, tastes, and environment in so far as they are independent of the choices made" (p. 1285).

We examine a deterministic discrete time infinite horizon optimal control problem. The initial capital stock is fixed: $k(0) \equiv k_0$. The evolution equation of the state $k(t)$ at date $t$ is

(1) $$k(t+1) = f_{t+1}(k(t), c(t)), \text{ with } c(t) \in \mathrm{K}_t(k(t)).$$

The set $K_t(k(t))$ of possible decisions $c(t)$ at date $t$ depends on $t$ and on the current state $k(t)$. There is an infinite sequence of gains

(2) $\qquad U_t(c(t), k(t)), \ t \geq 0,$

the sum of which may be unbounded, and is to be maximized under two commonly used optimality criteria to be defined below. $D_t$ is the set of states at date $t$ from which future evolutions are defined: for all $k(t)$ in $D_t$, the set $K_t(k(t))$ is nonempty, and for all $c(t)$ in $K_t(k(t))$, $U_t(c(t), k(t))$ and $f_{t+1}(c(t), k(t))$ are defined with values respectively in the real set $\mathbb{R}$ and $D_{t+1}$. We define an attainable path as an infinite sequence that starts at date 0 from $k(0) \equiv k_0$, with the state and decision being $(c(t), k(t))$ such that $c(t) \in K_t(k(t))$ and $k(t+1) = f_{t+1}(c(t), k(t)) \in D_{t+1}$, for all $t \geq 0$.

## 2.2 The Overtaking Criterion

We explicitly incorporate the boundary condition that "no goods should be wasted" and convert the attainable paths by changing the values of both consumption and capital at a newly chosen "terminal date", $T$, so that both are 0 at $T+1$:

(3) $\qquad \text{for all } T, \ \tilde{c}(t) = \begin{cases} c(t), 0 < t < T-1, \\ \tilde{c}(T), \text{ s.t. } f_T(\tilde{c}(T), k(T)) = 0, t = T, \\ 0, t \geq T+1, \end{cases}$

$$\tilde{k}(t) = \begin{cases} k(t), 0 < t \leq T, \\ 0, t \geq T+1. \end{cases}$$

The image of this conversion is as follows: the "doomsday", $T$, suddenly arrives, and the agents are able to "eat up" the entire stocks of goods at short notice before the end of the world and therefore, nothing is wasted. For two such converted attainable paths $(k_1, c_1)$ and $(k_2, c_2)$, we define "<" (i.e., $(c_2, k_2)$ overtakes $(c_1, k_1)$) as follows:

**Definition 1.** For $U_t \geq 0$, all $t$, $(k_1, c_1) < (k_2, c_2)$ if

$$\lim_{T \to \infty} \left( \left( \sum_{t=0}^{T-1} (U_t(c_2(t), k_2(t))) + U_T(\tilde{c}_2(T), k_2(T)) \right) - \left( \sum_{t=0}^{T-1} (U_t(c_1(t), k_1(t))) + U_T(\tilde{c}_1(T), k_1(T)) \right) \right) > 0.$$

As in Cai and Nitta (2007), based on Brock's (1970) notion of weak maximality, our optimality criterion is defined as follows: an attainable path $(\bar{k}, \bar{c})$ is optimal if no other attainable path $(k, c)$ overtakes it. Again, we consider the case $U_t(\bar{c}(t), \bar{k}(t)) \geq 0$ and $U_t(c(t), k(t)) \geq 0$, all $t$.[3] We define the optimality criterion as: no other attainable path $(k_1, c_1)$ overtakes $(\bar{k}, \bar{c})$ if

$$\lim_{T \to \infty} \left( \left( \sum_{t=0}^{T-1} (U_t(c(t), k(t))) + U_T(\tilde{c}(T), k(T)) \right) - \left( \sum_{t=0}^{T-1} (U_t(\bar{c}(t), \bar{k}(t))) + U_T(\tilde{\bar{c}}(T), \bar{k}(T)) \right) \right) \leq 0.$$

Denote $(c_T, k_T)$ to be the optimal solution for the maximization problem of the finite horizon ($T$) version of (2), subject to (1) and $k(T+1) = 0$; and denote $(c^\circ, k^\circ)$ to be its limit as the horizon $T$ grows to infinity, i.e., $c^\circ(t) \equiv \lim_{T \to \infty} c_T(t)$, $k^\circ(t) \equiv \lim_{T \to \infty} k_T(t)$.

---

[3] The case $U_t(\bar{c}(t), \bar{k}(t)) \leq 0$ and $U_t(c(t), k(t)) \leq 0$, all $t$, can be similarly defined.

The following theorem generalizes Theorem 1 of Cai and Nitta (2007). The proof of the theorem, however, wound need the following lemma:

**Lemma 1.** Let $a_T$ and $b_T$, $T \in [0,\infty)$, be two sequences. If $\varliminf_{T\to\infty} a_T > 0$, $\varliminf_{T\to\infty} b_T > 0$, then $\varliminf_{T\to\infty}(ab)_T = \varliminf_{T\to\infty} a_T \cdot \varliminf_{T\to\infty} b_T$.

**Proof.** See the proof for Lemma 1 in Cai and Nitta (2007).

**Theorem.** If

$$\lim_{T\to\infty}\left\{\frac{\sum_{t=0}^{T-1}\left\{\left(U_t\left(c_T(t),k_T(t)\right)-U_t\left(c^\circ(t),k^\circ(t)\right)\right)\right\}+\left\{U_T\left(c_T(T),k_T(T)\right)-U_T\left(\tilde{c}^\circ(T),k^\circ(T)\right)\right\}}{\sum_{t=0}^{T}\left\{U_t\left(c_T(t),k_T(t)\right)\right\}}\right\}=0,$$

then no other attainable path $(k_1,c_1)$ overtakes $(k^\circ,c^\circ)$.

**Proof:** See Appendix A.

We have thus identified the condition under which the limit of the solutions for the finite horizon problems is optimal among all attainable paths for the infinite horizon optimization problems in $\mathbb{R}$, in the sense of the overtaking criterion, for the general problem. Specifically, as in Cai and Nitta (2007), the condition in Theorem requires that the loss accompanying the limit practice is negligible as compared to the value of the maximand.

## 2.3  The Sum-of-Utilities Criterion

We proceed to consider another commonly adopted welfare criterion: the sum-of-utilities, which is stronger than the overtaking criterion. Definition 2 and 3 define the unique optimum under such a criterion.

**Definition 2.** For any attainable paths $(\hat{k},\hat{c})$ converted under formula (3), if

$$\lim_{T\to\infty}\left(\left(\sum_{t=0}^{T-1}\left(U_t(\hat{c},\hat{k})-U_t(c,k)\right)\right)+\left(U_T(\tilde{\hat{c}}(T),\hat{k}(T))-U_T(\tilde{c}(T),k(T))\right)\right)\leq 0, \text{ then } (k,c)$$

is the unique optimum.

**Definition 3.** For any attainable paths $(\hat{k},\hat{c})$ converted under formula (3), if

$$\lim_{T\to\infty}\left(\sum_{t=0}^{T}\left(U_t(\hat{c},\hat{k})-U_t(c,k)\right)\right)\leq 0, \text{ then } (k,c) \text{ is the unique optimum.}$$

**Remark.** Definition 2 and 3 are equivalent when

$\lim_{T\to\infty}U_T(\tilde{\hat{c}}(T),\hat{k}(T))-\lim_{T\to\infty}U_T(\tilde{c}(T),k(T))=0$. Moreover, Definition 1 and 2 are

equivalent when $\lim_{T\to\infty}\left(\left(\sum_{t=0}^{T-1}\left(U_t(\hat{c},\hat{k})-U_t(c^\circ,k^\circ)\right)\right)+\left(U_T(\tilde{\hat{c}}(T),\hat{k}(T))-U_T(\tilde{c}^\circ(T),k^\circ(T))\right)\right)$

exists, as in that case "lim inf" can be replaced by "lim".

We have thus identified the conditions under which the derived path is unique optimal, in the sense of the sum-of-utilities criterion. Note that for the undiscounted case, the conditions may not be satisfied as

$$\lim_{T\to\infty}\left(\left(\sum_{t=0}^{T-1}\left(U_t(\hat{c},\hat{k})-U_t(c,k)\right)\right)+\left(U_T(\tilde{\hat{c}}(T),\hat{k}(T))-U_T(\tilde{c}(T),k(T))\right)\right) \text{ may not exist, as}$$

shown in Cai and Nitta (2007).

## 3. Application: A One-Sector Growth Model

In this section, we consider the applicability of the results by examining a parametric example of the one-sector growth model. The properties of the optimal path are well-known for the case when the discount factor, $\beta$, is strictly less than 1.[4] On the other hand, Cai and Nitta (2007) consider the undiscounted case. The development here will completely parallel that in Cai and Nitta (2007). However, our focus here is to consider the impacts of discounting on the optimal capital accumulation paths. We consider a social planner's problem, in which the discount factor $\beta$ is not restricted to $(0,1]$. The planner's objective is

(4) $$\max_{c(t)} \sum_{t=0}^{T} \beta^t \ln(c(t)), \text{ where } T \in [0,\infty),$$

subject to

(5) $$c(t) + k(t+1) = f(k(t)), \text{ where } f(k(t)) = k(t)^\alpha,$$

$$0 < \alpha < 1, \quad \beta \in (0, 1/\alpha), \quad k(0) \text{ given,}$$

$$0 < k(t) < 1, \quad k(T+1) = 0.$$

Here we assume that $\beta \in (0, 1/\alpha)$ to ensure a positive consumption level over time. As $0 < k(t) < 1$, $0 < c(t) < 1$. The paths of $c(t)$ and $k(t)$ are uniquely determined given $k(0) > 0$ and $k(T+1) = 0$, with $k(t+1) = \alpha\beta \dfrac{1-(\alpha\beta)^{T-t}}{1-(\alpha\beta)^{T-t+1}} k^\alpha(t)$, $t = 0,1,\ldots,T$.

---
[4] See, for example, Stocky and Lucas (1989).

The optimal solution $(k_T(t), c_T(t))$ for time horizon $T$ is

$$k_T(t) = \frac{\left(1-(\alpha\beta)^{T-t+1}\right)(\alpha\beta)^{\frac{1-\alpha^t}{1-\alpha}} k(0)^{\alpha^t}}{\left(1-(\alpha\beta)^{T-t+2}\right)^{1-\alpha}\left(1-(\alpha\beta)^{T-t+3}\right)^{\alpha(1-\alpha)} \cdots \left(1-(\alpha\beta)^{T+1}\right)^{\alpha^{t-1}}},$$

$$c_T(t) = \frac{\left(1-(\alpha\beta)\right)(\alpha\beta)^{\frac{\alpha(1-\alpha^t)}{1-\alpha}} k(0)^{\alpha^{t+1}}}{\left(1-(\alpha\beta)^{T-t+1}\right)^{1-\alpha}\left(1-(\alpha\beta)^{T-t+2}\right)^{\alpha(1-\alpha)} \cdots \left(1-(\alpha\beta)^{T}\right)^{\alpha^{t-1}(1-\alpha)}\left(1-(\alpha\beta)^{T+1}\right)^{\alpha^t}}.$$

When time approaches to infinity, their limits are

$$k^\circ(t) \equiv \lim_{T\to\infty} k_T(t) = (\alpha\beta)^{\alpha^{t-1}+\cdots+1} K_0^{\alpha^t} = (\alpha\beta)^{1/(1-\alpha)} \left(\frac{k(0)}{(\alpha\beta)^{1/(1-\alpha)}}\right)^{\alpha^t},$$

$$c^\circ(t) \equiv \lim_{T\to\infty} c_T(t) = (1-\alpha\beta)(\alpha\beta)^{\alpha/(1-\alpha)} \left(\frac{k(0)}{(\alpha\beta)^{1/(1-\alpha)}}\right)^{\alpha^{t+1}},$$

$$\lambda^\circ(t) \equiv \lim_{T\to\infty} \lambda_T(t) = (1-\alpha\beta)^{-1} (\alpha\beta)^{-\alpha/(1-\alpha)} \left(\frac{k(0)}{(\alpha\beta)^{1/(1-\alpha)}}\right)^{-\alpha^{t+1}}.$$

In Appendix C and D, we show that the conditions in Theorem and Definition 2 and 3 are satisfied and $(c^\circ, k^\circ)$ is optimal among all the attainable paths in the sense of the overtaking criterion, as well as unique optimal under the sum-of-utilities criterion. We immediately have:

**Result 1**. *The optimal saving ratio $r$ is constant over time:* $r \equiv \dfrac{k°(t+1)}{f(k°(t))} = \alpha\beta$.

Cai and Nitta (2007) show that the optimal saving ratio depends exclusively on the productivity $\alpha$. Here we show that it also depends on the discount factor $\beta$, and the lower is the discount factor, the less is the optimal saving ratio.

**Proposition 1.** (Convergence) $c°(t)$, $k°(t)$, and $\lambda°(t)$ converge to finite positive values as $t \to \infty$: $c_\infty \equiv \lim_{t\to\infty} c°(t) = (1-\alpha\beta)(\alpha\beta)^{\frac{\alpha}{1-\alpha}} > 0$, $k_\infty \equiv \lim_{t\to\infty} k°(t) = (\alpha\beta)^{\frac{1}{1-\alpha}} > 0$, $\lambda_\infty \equiv \lim_{t\to\infty} \lambda°(t) = (1-\alpha\beta)^{-1}(\alpha\beta)^{\frac{-\alpha}{1-\alpha}} > 0$. When $k(0) > (\alpha\beta)^{\frac{1}{1-\alpha}}$ $\left(k(0) < (\alpha\beta)^{\frac{1}{1-\alpha}}\right)$, $k°(t)$ and $c°(t)$ are monotonically decreasing (increasing) in $t$, whereas $\lambda°(t)$ is monotonically increasing (decreasing) in $t$. $k°(t)$, $c°(t)$, and $\lambda°(t)$ are in steady states when $k(0) = (\alpha\beta)^{\frac{1}{1-\alpha}}$. Moreover, the speed of convergence slows down as the economy approaches the steady state.

**Proof.** See Appendix B.

Proposition 1 indicates that the economy converges to a steady state. We see that compared to the case of $\beta = 1$, applying a positive discount factor $\beta \in (0,1)$ results in a fall in consumption levels in and out of steady states. Formally, at time $t$, the ratio between the consumption levels in the two cases is $(\beta)^{\frac{1-\alpha^t}{1-\alpha}} < 1$, which falls over time. Moreover, at a given time, given the value of $\alpha$, the less the future value is discounted, the higher will be the ratio. Finally, as the capital stock converges to a positive value, it follows that $\lim_{t\to\infty}[\lambda(t)] \neq 0$; therefore, the usual transversality condition is violated.

However, as in Appendix E, it can be shown that the problem satisfies the boundary condition that is a natural extension of the finite horizon boundary conditions at $T = \infty$.

**Proposition 2.** (The productivity effect) The effect of the elasticity of output with respect to capital, $\alpha$, dominates the effect of the initial capital value $k(0)$ as $t \to \infty$. In the steady state, both $k_\infty$ and $c_\infty$ are decreasing in the $\alpha$. Moreover, when $\alpha \to 1$, $k_\infty \to 0$ if $\beta \in (0,1)$; $k_\infty \to e^{-1}$ if $\beta = 1$; and $k_\infty \to \infty$ if $\beta > 1$.

Proposition 2 shows that the higher the level of the productivity, the higher are the stationary values in the steady state. Moreover, we see that at extremely low levels of productivity, when $t \to \infty$, capital approaches zero at a positive discount rate, a positive value when the rate is zero, and infinity at a negative rate. In other words, the effects of discounting may be extremely consequential when the level of productivity is relatively low.

Moreover, it can be easily verified that the other properties reported in Cai and Nitta (2007), such as the property of path dependency (Proposition 3 in Cai and Nitta (2007)), and the dependence of capital, consumption, and shadow price on the length of the planning horizon (Proposition 4 in Cai and Nitta (2007)) largely hold. For most other parametric examples, however, it may not be possible to explicitly check the conditions in Theorem and Definition 2 and 3 and study the resultant paths. In such cases, we can resort to a numerical approach to compute explicit solutions and to check the conditions.

## 4. Discussion: The Legitimacy of the Approach

By resorting to the arguments in the non-standard analysis, Cai and Nitta (2007) extend real numbers to hyper-real numbers and present a mathematically correct infinite horizon extension of the finite horizon problem. They demonstrate that there exists a unique optimal solution to the infinite horizon problem in the hyper-real number field, with the "projection" of which on the real number field being the limit of the solutions to the undiscounted finite horizon problems. As shown in Appendix E, the hyper-real extension for the general problem can be similarly developed, and we can establish the legitimacy of the derived optimal paths by showing that they are the standard parts of the unique optimal paths in the hyper-real number field that satisfies the boundary condition that goods will not be wasted at the infinite terminal time.

## 5. Conclusions

In this paper, we present a generalization to Cai and Nitta's (2007) approach to find optimums for problems in infinite-horizon dynamics with boundary conditions at infinite time. Our generalization improves the applicability of their approach. Nevertheless, for a systematic re-examination of infinite horizon optimization problems, one may still need to consider continuous time, and other criteria of intertemporal optimality.

# APPENDIX A: PROOF OF THEOREM

For an arbitrary number $T$, as $c_T$ is the optimal solution, we have

$$\frac{\sum_{t=0}^{T-1} U_t\left(c_1(t), k_1(t)\right) + U_T\left(c_1(T), k_1(T)\right)}{\sum_{t=0}^{T} U_t\left(c_T(t), k_T(t)\right)} \leq 1.$$

Also, as

$$\frac{\sum_{t=0}^{T-1} U_t\left(c_1(t), k_1(t)\right) + U_T\left(c_1(T), k_1(T)\right)}{\sum_{t=0}^{T-1} U_t\left(c^\circ(t), k^\circ(t)\right) + U_T\left(\tilde{c}^\circ(T), k^\circ(T)\right)} =$$

$$\left( \frac{\sum_{t=0}^{T-1} U_t\left(c_1(t), k_1(t)\right) + U_T\left(c_1(T), k_1(T)\right)}{\sum_{t=0}^{T} U_t\left(c_T(t), k_T(t)\right)} \right) \cdot \left( \frac{\sum_{t=0}^{T} U_t\left(c_T(t), k_T(t)\right)}{\sum_{t=0}^{T-1} U_t\left(c^\circ(t), k^\circ(t)\right) + U_T\left(\tilde{c}^\circ(T), k^\circ(T)\right)} \right).$$

Therefore,

$$\frac{\sum_{t=0}^{T} U_t\left(c_T(t), k_T(t)\right)}{\sum_{t=0}^{T-1} U_t\left(c^\circ(t), k^\circ(t)\right) + U_T\left(\tilde{c}^\circ(T), k^\circ(T)\right)}$$

$$= \frac{\sum_{t=0}^{T} U_t\left(c_T(t), k_T(t)\right)}{\sum_{t=0}^{T} U_t\left(c_T(t), k_T(t)\right) - \left( \sum_{t=0}^{T-1} U_t\left(c_T(t), k_T(t)\right) - \sum_{t=0}^{T-1} U_t\left(c^\circ(t), k^\circ(t)\right) \right) - U_T\left(c_T(T), k_T(T)\right) + U_T\left(\tilde{c}^\circ(T), k^\circ(T)\right)}$$

$$= \frac{1}{1 - \frac{\left(\sum_{t=0}^{T-1} U_t(c_T(t), k_T(t)) - \sum_{t=0}^{T-1} U_t(c^\circ(t), k^\circ(t))\right) + U_T(c_T(T), k_T(T)) - U_T(\tilde{c}^\circ(T), k^\circ(T))}{\sum_{t=0}^{T} U_t(c_T(t), k_T(t))}}.$$

From the condition in Theorem, we see that $\displaystyle\lim_{T \to \infty} \frac{\sum_{t=0}^{T} U_t(c_T(t), k_T(t))}{\sum_{t=0}^{T-1} U_t(c^\circ(t), k^\circ(t)) + U_T(\tilde{c}^\circ(T), k^\circ(T))} = 1$.

Therefore, from Lemma 1, we have

$$\lim_{T \to \infty} \frac{\sum_{t=0}^{T-1} U_t(c_T(t), k_T(t)) + U_T(c_T(T), k_T(T))}{\sum_{t=0}^{T-1} U_t(c^\circ(t), k^\circ(t)) + U_T(\tilde{c}^\circ(T), k^\circ(T))}$$

$$= \lim_{T \to \infty} \left( \frac{\sum_{t=0}^{T-1} U_t(c_T(t), k_T(t)) + U_T(c_T(T), k_T(T))}{\sum_{t=0}^{T} U_t(c_T(t), k_T(t))} \right) \cdot \lim_{T \to \infty} \left( \frac{\sum_{t=0}^{T} U_t(c_T(t), k_T(t))}{\sum_{t=0}^{T-1} U_t(c^\circ(t), k^\circ(t)) + U_T(\tilde{c}^\circ(T), k^\circ(T))} \right) \leq 1.$$

One the other hand, again from Lemma 1, we see that

$$\lim_{T \to \infty} \left( \left( \sum_{t=0}^{T-1} (U_t(c(t), k(t))) + U_T(\tilde{c}(T), k(T)) \right) - \left( \sum_{t=0}^{T-1} (U_t(\bar{c}(t), \bar{k}(t))) + U_T(\tilde{\bar{c}}(T), \bar{k}(T)) \right) \right)$$

$$= \lim_{T \to \infty} \left( 1 - \frac{\sum_{t=0}^{T-1} (U_t(c(t), k(t))) + U_T(\tilde{c}(T), k(T))}{\sum_{t=0}^{T-1} (U_t(\bar{c}(t), \bar{k}(t))) + U_T(\tilde{\bar{c}}(T), \bar{k}(T))} \right) \cdot \lim_{T \to \infty} \left( \sum_{t=0}^{T-1} (U_t(\bar{c}(t), \bar{k}(t))) + U_T(\tilde{\bar{c}}(T), \bar{k}(T)) \right).$$

Because $\displaystyle\lim_{T \to \infty} \left( \sum_{t=0}^{T-1} (U_t(\bar{c}(t), \bar{k}(t))) + U_T(\tilde{\bar{c}}(T), \bar{k}(T)) \right) > 0$ and

$\displaystyle\lim_{T \to \infty} \frac{\sum_{t=0}^{T-1} U_t(c_T(t), k_T(t)) + U_T(c_T(T), k_T(T))}{\sum_{t=0}^{T-1} U_t(c^\circ(t), k^\circ(t)) + U_T(\tilde{c}^\circ(T), k^\circ(T))} \leq 1$, we see that

$$\lim_{T \to \infty}\left[\left(\sum_{t=0}^{T-1}\left(U_t\left(c(t),k(t)\right)\right)+U_T\left(\tilde{c}(T),k(T)\right)\right)-\left(\sum_{t=0}^{T-1}\left(U_t\left(c^\circ(t),k^\circ(t)\right)\right)+U_T\left(\tilde{c}^\circ(T),k^\circ(T)\right)\right)\right] \leq 0.$$

<p align="right">Q.E.D.</p>

## APPENDIX B: PROOF OF PROPOSITION 1:

$$\frac{\dot{c}^\circ(t)}{c^\circ(t)} = \frac{d\ln c^\circ(t)}{dt} = \ln\alpha \ln\left(\frac{k(0)}{(\alpha\beta)^{\frac{1}{1-\alpha}}}\right)\cdot\alpha^{t+1}, \quad \frac{\dot{k}^\circ(t)}{k^\circ(t)} = \frac{d\ln k^\circ(t)}{dt} = \ln\alpha \ln\left(\frac{k(0)}{(\alpha\beta)^{\frac{1}{1-\alpha}}}\right)\cdot\alpha^{t}.$$

<p align="right">Q.E.D.</p>

## APPENDIX C: PROOF OF THE FACT THAT THE EXAMPLE SATISFIES THE CONDITION IN THEOREM

We first show that for a finite $t$, for $\beta \in (0,1)$, $\ln(c_T(t)) - \ln c^\circ(t)$ is 0 when $T \to \infty$. We see that

$$\ln(c_T(t)) - \ln c^\circ(t) = -(1-\alpha)\alpha^{-(T-t+1)}\left\{\alpha^{T-t+1}\ln\left(1-(\alpha\beta)^{T-t+1}\right)+\cdots+\alpha^T\ln\left(1-(\alpha\beta)^T\right)\right\} - \alpha^t\ln\left(1-(\alpha\beta)^{T+1}\right),$$

(A1)

when $T-t+1 \leq s \leq T$, the right-hand side of (A1) is

$$\text{RHS of (A1)} = (1-\alpha)\alpha^{-(T-t+1)}\sum_{s=T-t+1}^{T}\alpha^s\left(-\ln\left(1-(\alpha\beta)^s\right)\right)+\alpha^t\left(-\ln\left(1-(\alpha\beta)^{T+1}\right)\right). \quad \text{(A2)}$$

We show that the right-hand side of (A2) approaches zero when $t \to \infty$.

We observe that because

$$-\ln\left(1-(\alpha\beta)^s\right) = (\alpha\beta)^s + \frac{1}{2}(\alpha\beta)^{2s} + \cdots < (\alpha)^s + \frac{1}{2}(\alpha)^{2s} + \cdots = -\ln\left(1-\alpha^s\right),$$

$0 < -\ln\left(1-(\alpha\beta)^s\right) < -\ln\left(1-\alpha^s\right)$, (A2) is bounded by

$$(1-\alpha)\alpha^{-(T-t+1)}\sum_{s=T-t+1}^{T}\alpha^s\left(-\ln\left(1-(\alpha)^s\right)\right) + \alpha^t\left(-\ln\left(1-(\alpha)^{T+1}\right)\right).$$

Accordingly to Cai and Nitta (2007),

$$\lim_{T\to\infty}\sum_{t=0}^{T}\left((1-\alpha)\alpha^{-(T-t+1)}\sum_{s=T-t+1}^{T}\alpha^s\left(-\ln\left(1-(\alpha)^s\right)\right) + \alpha^t\left(-\ln\left(1-(\alpha)^{T+1}\right)\right)\right) \text{ is a finite}$$

positive number. Therefore, $\lim_{T\to\infty}\sum_{t=0}^{T}\left(\ln c_T(t) - \ln c^\circ(t)\right)$ is also a finite positive number.

Let $M = \lim_{T\to\infty}\sum_{t=0}^{T}\left(\ln c_T(t) - \ln c^\circ(t)\right) > 0$.

We proceed to show that $\lim_{T\to\infty}\sum_{t=0}^{T-1}\beta^t\left(\ln c_T(t) - \ln c^\circ(t)\right) = 0$. For all $t > 0$, there

exists a $T_1 > 0$, such that

$$\sum_{t=0}^{T-1}\beta^t\left(\ln(c_T(t)) - \ln c^\circ(t)\right) = \sum_{t=0}^{T_1-1}\beta^t\left(\ln(c_T(t)) - \ln c^\circ(t)\right) + \sum_{t=T_1}^{T-1}\beta^t\left(\ln(c_T(t)) - \ln c^\circ(t)\right)$$

$$< \sum_{t=0}^{T_1-1}\beta^t\left(\ln(c_T(t)) - \ln c^\circ(t)\right) + \sum_{t=T_1}^{T-1}\beta^{T_1}\left(\ln(c_T(t)) - \ln c^\circ(t)\right)$$

$$= \sum_{t=0}^{T_1-1}\beta^t\left(\ln(c_T(t)) - \ln c^\circ(t)\right) + \beta^{T_1}\sum_{t=T_1}^{T-1}\left(\ln(c_T(t)) - \ln c^\circ(t)\right).$$

For an arbitrary $T$, we have $0 < \sum_{t=0}^{T}\left(\ln c_T(t) - \ln c^\circ(t)\right) < M$. $\forall \varepsilon > 0$, $\exists T_1$, we fix a $T_1$ such that $\beta^{T_1} < \dfrac{\varepsilon}{2M}$. Therefore,

$$\sum_{t=0}^{T-1} \beta^t \left(\ln(c_T(t)) - \ln c^\circ(t)\right) = \sum_{t=0}^{T_1-1} \beta^t \left(\ln(c_T(t)) - \ln c^\circ(t)\right) + \sum_{t=T_1}^{T-1} \beta^t \left(\ln(c_T(t)) - \ln c^\circ(t)\right)$$

$$< \sum_{t=0}^{T_1-1} \beta^t \left(\ln(c_T(t)) - \ln c^\circ(t)\right) + \beta^{T_1} M.$$

As for all $t$, $\lim_{T \to \infty} c_T(t) = c^\circ(t)$. For $0 \leq t \leq T_1 - 1$, $\exists T_0$ such that when $T > T_0$,

$$\left(\ln(c_T(t)) - \ln c(t)\right) < \dfrac{\varepsilon}{2T_1}, \text{ for } 0 \leq t \leq T_1 - 1.$$

When $T > T_0$,

$$\sum_{t=0}^{T-1} \beta^t \left(\ln(c_T(t)) - \ln c^\circ(t)\right) = \sum_{t=0}^{T_1-1} \beta^t \left(\ln(c_T(t)) - \ln c^\circ(t)\right) + \sum_{t=T_1}^{T-1} \beta^t \left(\ln(c_T(t)) - \ln c^\circ(t)\right)$$

$$< \sum_{t=0}^{T_1-1} \beta^t \left(\ln(c_T(t)) - \ln c^\circ(t)\right) + \beta^{T_1} M$$

$$< T_1 \dfrac{\varepsilon}{2T_1} + \beta^{T_1} M$$

$$< T_1 \dfrac{\varepsilon}{2T_1} + \dfrac{\varepsilon}{M} M = \varepsilon.$$

Therefore, we have shown that $\lim_{T \to \infty} \sum_{t=0}^{T-1} \beta^t \left(\ln c_T(t) - \ln c^\circ(t)\right) = 0$.

Finally, we consider $\ln c_T(T) - \ln c^\circ(T)$. As $f(k_T(T)) = c_T(T)$,

$$\ln c_T(T) - \ln c^\circ(T) = (1-\alpha)\alpha^{-1}\sum_{s=1}^{T}\alpha^s\left(-\ln\left(1-(\alpha\beta)^s\right)\right) + \alpha^T\left(-\ln\left(1-(\alpha\beta)^{T+1}\right)\right).$$

We know $\lim_{T\to\infty}\alpha^T\left(-\ln\left(1-(\alpha\beta)^{T+1}\right)\right) = 0$, and because

$$\sum_{s=1}^{\infty}\alpha^s\left(-\ln\left(1-(\alpha\beta)^s\right)\right) < \sum_{s=1}^{\infty}\alpha^s\left(-\ln\left(1-(\alpha)^s\right)\right),$$

let $\alpha^s = x$, $1-x = y$, we see that

$$\sum_{s=1}^{T}\alpha^s\left(-\ln\left(1-(\alpha)^s\right)\right) < \frac{1}{\ln\alpha}\int_{1-\alpha}^{1-\alpha^T}\ln y\, dy = \frac{1}{\ln\alpha}\left[y\ln y - y\right]_{1-\alpha}^{1-\alpha^T}$$

$$= \frac{1}{\ln\alpha}\left((1-\alpha^T)\ln(1-\alpha^T) - (1-\alpha)\ln(1-\alpha) - (1-\alpha^T) + (1-\alpha)\right),$$

$$< \frac{1}{\ln\alpha}\left(-(1-\alpha)\ln(1-\alpha) + (1-\alpha)\right),$$

which is a finite number.

Therefore, we see that $\beta^T\left(\ln c_T(T) - \ln c^\circ(T)\right)$ approaches 0 when $T$ approaches infinity.

*Q.E.D.*

## APPENDIX D: PROOF OF THE FACT THAT THE EXAMPLE SATISFIES THE CONDITIONS IN DEFINITION 2 and 3:

When $\beta \in (0,1)$, $\lim_{T\to\infty}\beta^T U\left(k^\circ(T)^\alpha\right) = \lim_{T\to\infty}\beta^T \ln\left((\alpha\beta)^{\frac{1}{1-\alpha}}\left(\frac{k(0)}{(\alpha\beta)^{1/(1-\alpha)}}\right)^{\alpha^t}\right) = 0$.

Therefore, when $\lim_{T\to\infty} \beta^T U\left(\hat{k}(T)^\alpha\right) = 0$, we see that Definition 2 and 3 are equivalent for the example in Section 3.1. Note that $\lim_{T\to\infty} \beta^T U\left(\hat{k}(T)^\alpha\right) = 0$ is generally satisfied, as for an exponential function $\hat{k}(T) = a^T$, $a \in \mathbb{R}_+$, $\lim_{T\to\infty} \beta^T \ln\left(a^T\right) = \lim_{T\to\infty} T\beta^T \alpha \ln(\alpha) = 0$.

We consider Definition 2. We see that

$$\lim_{T\to\infty}\left(\sum_{t=0}^{T-1}\left(U_t\left(\hat{c},\hat{k}\right) - U_t\left(c^\circ,k^\circ\right)\right) + U_T\left(\tilde{c}(T),\hat{k}(T)\right) - U_T\left(\tilde{c}^\circ(T),k^\circ(T)\right)\right)$$

$$= \lim_{T\to\infty}\left(\underbrace{\left(\sum_{t=0}^{T-1}\left(U_t\left(c^\circ,k^\circ\right)\right) + U_T\left(\tilde{c}^\circ(T),k^\circ(T)\right)\right)}_{\{I\}} \left(\underbrace{\frac{\sum_{t=0}^{T-1}\left(U_t\left(\hat{c},\hat{k}\right)\right) + U_T\left(\tilde{c}(T),\hat{k}(T)\right)}{\sum_{t=0}^{T-1}\left(U_t\left(c^\circ,k^\circ\right)\right) + U_T\left(\tilde{c}^\circ(T),k^\circ(T)\right)} - 1}_{\{II\}}\right)\right).$$

For our specific model, we first consider term $\{I\}$. As

$$\sum_{t=0}^{T-1} \beta^t \ln\left((1-\alpha\beta)(\alpha\beta)^{\frac{\alpha}{1-\alpha}}\left(\frac{k(0)}{(\alpha\beta)^{1/(1-\alpha)}}\right)^{\alpha^{t+1}}\right) + \beta^T \ln\left((\alpha\beta)^{\frac{1}{1-\alpha}}\left(\frac{k(0)}{(\alpha\beta)^{1/(1-\alpha)}}\right)^{\alpha^t}\right)$$

$$= \sum_{t=0}^{T-1} \beta^t \left(\ln\left((1-\alpha\beta)(\alpha\beta)^{\frac{\alpha}{1-\alpha}}\right) + (\alpha\beta)^t \alpha \ln\left(\frac{k(0)}{(\alpha\beta)^{1/(1-\alpha)}}\right)\right) + \beta^T \ln\left((\alpha\beta)^{\frac{1}{1-\alpha}}\left(\frac{k(0)}{(\alpha\beta)^{1/(1-\alpha)}}\right)^{\alpha^t}\right),$$

we see it is a geometric series that converges when $T \to \infty$.

Next, we consider term $\{II\}$. Because we have shown in Appendix C that

$$\lim_{T\to\infty}\left(\sum_{t=0}^{T-1} \beta^t \ln\left(c_T(t)\right) + \beta^T \ln\left(k_T(T)^\alpha\right) - \sum_{t=0}^{T-1} \beta^t \ln\left(c^\circ(t)\right) - \beta^T \ln\left(k^\circ(T)^\alpha\right)\right) = 0,$$

we can replace $\left(c^\circ, k^\circ\right)$ with $\left(c_T, k_T\right)$. As $\left(c_T, k_T\right)$ is the optimal solution for the planning horizon $T$, we then have $\dfrac{\sum_{t=0}^{T-1}\beta^{t}\left(\ln\left(\hat{c}(t)\right)\right)+\beta^{T}\ln\left(\left(\hat{k}(T)\right)^{\alpha}\right)}{\sum_{t=0}^{T-1}\beta^{t}\left(\ln\left(c_{T}(t)\right)\right)+\beta^{T}\ln\left(\left(k_{T}(T)\right)^{\alpha}\right)}\geq 1$. This completes our check and we see that the example satisfies both Definition 2 and 3 with the derived path being unique optimal.

*Q.E.D.*

# APPENDIX E: SOLVING THE INFINITE HORIZON OPTIMIZATION PROBLEMS WITH THE NON-STANDARD APPROACH

Excellent introductions to the non-standard analysis and surveys on the economic applications of the non-standard analysis are available in Anderson (1991) and Rubio (2000, Appendix). Following Anderson (1991, 2150-2151), we extend real numbers into hyper-real numbers. We define hyper-real number as $^{*}\mathbb{R}\equiv\mathbb{R}^{\mathbb{N}}/u$, where $\mathbb{N}$ is a set of natural numbers, $u$ is a free ultrafilter that is a maximal filter containing the *Fréchet filter*. The elements of $^{*}\mathbb{R}$ are represented as sequences and are denoted as $[<a_n>]$, where $a_n \in \mathbb{R}$. Instead of $T$, we consider an infinite star finite number $\tilde{T}\equiv [<T_n>]$. We extend $c(t)$ to $C(t)\equiv [<C_n(t)>]$, and $k(t)$ to $K(t)\equiv [<K_n(t)>]$. Moreover, $U_t : \mathbb{R}_+ \to \mathbb{R}$ and $f_t : \mathbb{R}_+ \to \mathbb{R}_+$ are extended to $^{*}U_t : {^{*}\mathbb{R}_+} \to {^{*}\mathbb{R}}$ and $^{*}f_t : {^{*}\mathbb{R}_+} \to {^{*}\mathbb{R}_+}$, respectively. Accordingly, $\sum_{t=0}^{T} U_t\left(c(t),k(t)\right)$ is extended to $^{*}\sum_{[<t_n>]=0}^{\tilde{T}} {^{*}U_t}\left(C(t),K(t)\right)$, with its elements represented by $\left[<\sum_{t=0}^{T_n} U_t(C_n(t))>\right]$. We

also denote

$$[<a_n>] \geq [<b_n>] \text{ if } \{n \in \mathbb{N} : a_n \geq b_n\} \in u,$$

$$\text{for } [<a_n>], [<b_n>] \in {}^*\mathbb{R}.$$

Therefore, the maximization problem can be reformulated as

*(1)
$$\max {}^*\!\!\sum_{[<t_n>]=0}^{\tilde{T}} {}^*U_t(C(t), K(t)),$$

subject to

*(2)
$$K([<t_n>]+1) = {}^*f_{t+1}(K([<t_n>]), C([<t_n>])),$$

$$0 \leq K([<t_n>]+1) \leq {}^*f_{t+1}(K([<t_n>]), C([<t_n>])), \quad t_n = 0, 1, \ldots, T_n,$$

given $K(0) > 0$, and,

the fact that goods will not be wasted: $K(\tilde{T}+1) = 0$.

Following the arguments of the non-standard analysis, we see that there exists a unique optimal solution $\{C(t), K(t)\}_{t=0}^{\tilde{T}}$ to problem *(1), subject to *(2). This is so because that under our extension of $\mathbb{R}$ to ${}^*\mathbb{R}$, the relation between $T$ and $T+1$ has been preserved to $\tilde{T}$ and $\tilde{T}+1$. In other words, the set of sequences $\{K(t)\}_{t=0}^{\tilde{T}}$ satisfying *(2) is a *closed, *bounded, and *convex subset of ${}^*\mathbb{R}$, and the objective function *(1) is continuous and strictly concave. Therefore, there exists a unique solution $\{C(t), K(t)\}_{t=0}^{\tilde{T}}$ to *(1) in the hyper-real number field.

Denote $\widetilde{st}$ to be the standard mapping from $^*\mathbb{R} \to \mathbb{R} \cup \{\pm\infty\}$. $\widetilde{st}$ projects the unique optimal path in the hyper-real number field to the real number field. As is widely known in the non-standard analysis, if there exist $\lim_{T\to\infty} c_T(t)$ and $\lim_{T\to\infty} k_T(t)$, we then have

$$\widetilde{st}(C_{\tilde{T}}(t)) = \lim_{T\to\infty} c_T(t) \equiv c°(t), \widetilde{st}(K_{\tilde{T}}(t)) = \lim_{T\to\infty} k_T(t) \equiv k°(t), \text{ where } t \in \mathbb{R}.$$

Therefore, from the non-standard optimal conditions, we have specified a unique path $(\widetilde{st}(C_{\tilde{T}}(t)), \widetilde{st}(K_{\tilde{T}}(t)))$ in $\mathbb{R}$. As shown in Theorem, once a fairly general condition is satisfied, such a path is optimal in $\mathbb{R}$, in the sense of the overtaking criterion.

*Q.E.D.*